# CALCULATION OF THE TOTAL ENERGY OF A DIATOMIC MOLECULE IN THE FIRST ORDER OF PERTURBATION THEORY TAKING INTO ACCOUNT THE PAULI PRINCIPLE AND PLASMA OSCILLATIONS OF ATOMIC ELECTRONS

**Koshcheev V.P.**[1,*], **Shtanov Yu.N.**[2,**]

[1] *Moscow Aviation Institute (National Research University), Strela Branch, Moscow oblast, Zhukovskii, Russia.*

[2] *Industrial University of Tyumen, Surgut Branch, Surgut, Russia.*

[*]*e-mail: koshcheev1@yandex.ru, **e-mail: yuran1987@mail.ru*

In the first order of perturbation theory, the total energy of a diatomic molecule in the ground state is calculated taking into account the Pauli principle and plasma oscillations of atomic electrons. The Fourier component of the potential energy of interaction of an atom with an atom has the form of a polynomial of the fourth degree of the atomic form factor. Numerical calculation is performed for the atomic form factor in the approximation of hydrogen-like wave functions, which approximate the solution of the Hartree-Fock equation for an isolated atom. It is shown that taking into account the plasma oscillations of atomic electrons leads to a self-consistent system of equations, the numerical solution of which makes it possible to determine the elastic constant, that is, the value of the second derivative at the minimum of the potential energy of the molecule. The total energy for nitrogen and fluorine molecules is calculated.

**Keywords:** potential energy of interaction, Pauli principle, hydrogen-like wave functions, Hartree-Fock equation, plasma oscillations.

Quite often, the results of calculating the total energy of diatomic molecules are compared with the well-studied data on fluorine and nitrogen molecules [1]. A new approach to calculating the potential energy of diatomic molecules [2] satisfactorily describes the experimental results for atoms of noble gases, if the atomic form – factor of an isolated atom is chosen in the Moliere approximation [3]. In this publication, the total energy of a diatomic nitrogen and fluorine molecule will be constructed in the first order of perturbation theory for the ground state, taking into account the Pauli principle and plasma oscillations of atomic electrons.

## THEORY

A diatomic molecule will be described using the stationary Schrödinger equation

$$H\psi = E\psi . \tag{1}$$

We represent the Hamiltonian of equation (1) in the form

$$H = H^0 + U; \tag{2}$$

$$U=\frac{Z_1Z_2e^2}{|\mathbf{r}_1-\mathbf{r}_2|}+\sum_{j_1=1}^{Z_1}\sum_{j_2=1}^{Z_2}\frac{e^2}{|\mathbf{r}_1+\mathbf{r}_{1j_1}-\mathbf{r}_2-\mathbf{r}_{2j_2}|}-\sum_{j_1=1}^{Z_1}\frac{Z_2e^2}{|\mathbf{r}_1+\mathbf{r}_{1j_1}-\mathbf{r}_2|}-\sum_{j_2=1}^{Z_2}\frac{Z_1e^2}{|\mathbf{r}_1-\mathbf{r}_2-\mathbf{r}_{2j_2}|}, \qquad (3)$$

where $U$ – potential energy of interaction of two atoms; $\mathbf{r}_1$ and $\mathbf{r}_2$ – coordinates of the first and second atomic nucleus; $\mathbf{r}_1+\mathbf{r}_{1j_1}$ and $\mathbf{r}_2+\mathbf{r}_{2j_2}$ – coordinates $j_1$ and $j_2$ electrons of the first and second atom, respectively; $\mathbf{r}_1-\mathbf{r}_2=\mathbf{r}+\delta\mathbf{r}$; plasma oscillations of atomic electrons are described by the vector $\delta\mathbf{r}$.

The solution to Eq. (1) with Hamiltonian (2) will be sought using the perturbation theory

$$\psi=\psi^0+\psi^1+\ldots$$

$$E=E^0+E^1+\ldots$$

We will seek the potential energy (electronic terms) of a diatomic molecule in the first order of the perturbation theory

$$E^1=\left\langle\psi^0\middle|U\middle|\psi^0\right\rangle, \qquad (4)$$

where angle brackets $\langle\ldots\rangle$ were introduced by Dirac [4].

We represent the Hamiltonian $H^0$ in the form

$$H^0=H_1^0+H_2^0,$$

where $H_i^0$ – Hamiltonian of the $i$-th atom; $i$=1,2.

The solution to the Schrödinger equation

$$H^0\psi^0=E^0\psi^0,$$

will be sought in the form

$$\psi^0=\psi_1^0\psi_2^0,$$

$$E^0=E_1^0+E_2^0,$$

where the Schrödinger equation for the $i$-th isolated atom has the form

$$H_i^0\psi_i^0=E_i^0\psi_i^0, \qquad (5)$$

where $\psi_i^0=\psi_i^0\left(\mathbf{r}_{i1},\mathbf{r}_{i2},\ldots,\mathbf{r}_{iZ_i}\right)$.

It is known [5] that using the variational principle from the stationary Schrödinger equation (4), one can construct the Hartree-Fock equation if the wave functions of atomic electrons are written in the form of Slater's determinant. Hydrogen-like wave functions that approximate the solution of the Hartree-Fock equation for an isolated atom are presented in [6]. Fluctuations in the potential energy of interaction (3) are caused by quantum fluctuations experienced by atomic electrons. Averaging over the quantum fluctuations of the position of atomic electrons will be carried out using the method [7], which Bethe used to calculate the atomic form factor, and we will perform averaging over the plasma oscillations of atomic electrons over the square of the modulus of the wave function of the harmonic oscillator in the ground state. The corresponding means will be denoted $\left\langle \psi_1^0 \middle| U \middle| \psi_1^0 \right\rangle = \langle ... \rangle_{e1}$, $\left\langle \psi_2^0 \middle| U \middle| \psi_2^0 \right\rangle = \langle ... \rangle_{e2}$ and $\left\langle \psi^0 \middle| U \middle| \psi^0 \right\rangle_{pl} = \langle ... \rangle_{pl}$.

We expand the potential interaction energy (3) into the Fourier integral

$$U = \int \frac{d^3\mathbf{k}}{(2\pi)^3} \left( \frac{4\pi Z_1 Z_2 e^2}{k^2} \exp\left(i\mathbf{k}\left(\mathbf{r}_1 - \mathbf{r}_2\right)\right) + \frac{4\pi e^2}{k^2} \sum_{j_1=1}^{Z_1} \sum_{j_2=1}^{Z_2} \exp\left(i\mathbf{k}\left(\mathbf{r}_{1j_1} - \mathbf{r}_{2j_2}\right)\right) - \right.$$
$$\left. - \frac{4\pi Z_2 e^2}{k^2} \sum_{j_1=1}^{Z_1} \exp\left(i\mathbf{k}\left(\mathbf{r}_2 - \mathbf{r}_{1j_1}\right)\right) - \frac{4\pi Z_1 e^2}{k^2} \sum_{j_2=1}^{Z_2} \exp\left(i\mathbf{k}\left(\mathbf{r}_1 - \mathbf{r}_{2j_2}\right)\right) \right). \tag{6}$$

Let us average (2) over the square of the modulus of the wave function of the electrons of the first atom

$$\langle U \rangle_{e1} = \int \frac{d^3\mathbf{k}}{(2\pi)^3} \left( \frac{4\pi Z_1 Z_2 e^2}{k^2} \exp\left(i\mathbf{k}\left(\mathbf{r}_1 - \mathbf{r}_2\right)\right) + \frac{4\pi e^2}{k^2} \left\langle \sum_{j_1=1}^{Z_1} \sum_{j_2=1}^{Z_2} \exp\left(i\mathbf{k}\left(\mathbf{r}_{1j_1} - \mathbf{r}_{2j_2}\right)\right) \right\rangle_{e1} - \right.$$
$$\left. - \frac{4\pi Z_2 e^2}{k^2} \left\langle \sum_{j_1=1}^{Z_1} \exp\left(i\mathbf{k}\left(\mathbf{r}_2 - \mathbf{r}_{1j_1}\right)\right) \right\rangle_{e1} - \frac{4\pi Z_1 e^2}{k^2} \sum_{j_2=1}^{Z_2} \exp\left(i\mathbf{k}\left(\mathbf{r}_1 - \mathbf{r}_{2j_2}\right)\right) \right); \tag{7}$$

$$\left\langle \sum_{j_1=1}^{Z_1} \sum_{j_2=1}^{Z_2} \exp\left(i\mathbf{k}\left(\mathbf{r}_{1j_1} - \mathbf{r}_{2j_2}\right)\right) \right\rangle_{e1} = F_1(k) \sum_{j_2=1}^{Z_2} \exp\left(i\mathbf{k}\left(\mathbf{r}_1 - \mathbf{r}_{2j_2}\right)\right);$$

$$\left\langle \sum_{j_1=1}^{Z_1} \exp\left(i\mathbf{k}\left(\mathbf{r}_2 - \mathbf{r}_{1j_1}\right)\right) \right\rangle_{e1} = F_1(k) \exp\left(i\mathbf{k}\left(\mathbf{r}_1 - \mathbf{r}_2\right)\right),$$

where $F_1(k)$ – atomic form factor; $F_1(0) = Z_1$.

Let us average $\langle U \rangle_{e1}$ over the square of the modulus of the wave function of the electrons of the second atom

$$\langle U \rangle_{e1,e2} = \int \frac{d^3\mathbf{k}}{(2\pi)^3} \left( \frac{4\pi Z_2 e^2}{k^2} \left(Z_1 - F_1(k)\right) \exp\left(i\mathbf{k}\left(\mathbf{r}_1 - \mathbf{r}_2\right)\right) - \frac{4\pi e^2}{k^2} \left(Z_1 - F_1(k)\right) \left\langle \sum_{j_2=1}^{Z_2} \exp\left(i\mathbf{k}\left(\mathbf{r}_1 - \mathbf{r}_{2j_2}\right)\right) \right\rangle_{e2} \right);$$

$$\left\langle \sum_{j_2=1}^{Z_2} \exp\left(i\mathbf{k}\left(\mathbf{r}_1 - \mathbf{r}_{2j_2}\right)\right) \right\rangle_{e2} = F_2(\mathrm{k}) \exp\left(i\mathbf{k}\left(\mathbf{r}_1 - \mathbf{r}_2\right)\right). \tag{8}$$

Let us average the potential energy of interaction of two atoms over the square of the modulus of the wave function of a harmonic oscillator in the ground state.

$$\langle U(r) \rangle_{e1,e2,\mathrm{pl.}} = \int \frac{d^3\mathbf{k}}{(2\pi)^3} \frac{4\pi e^2}{k^2} \left[Z_1 - F_1(k)\right]\left[Z_2 - F_2(k)\right] \exp\left[-\mathrm{k}^2 \sigma^2\right] \exp\left(i\mathbf{k}\mathbf{r}\right), \tag{9}$$

where $U(k,\sigma)=\frac{4\pi e^2}{k^2}\left[Z_1-F_1(k)\right]\left[Z_2-F_2(k)\right]\exp\left[-\mathrm{k}^2\sigma\right]$ – Fourier component of the potential energy of interaction of two atoms; $\sigma^2=\frac{\hbar}{4\omega m}$ – mean square of the amplitude of plasma oscillations of atomic electrons per degree of freedom; $m=m_e/2$; $m_e$ – electron mass; elastic constant $\omega^2 m=U''(\mathrm{r}_{\min})$ is the value of the second derivative at the minimum of the potential energy of the molecule масса электрона; $r=|\mathbf{r}|$ - distance between atoms in a molecule.

Similarly to how it is done in the kinetic theory [8], we add to the expression for the Fourier component of the potential energy of interaction of two atoms the factor $(1-F(k)/Z)$, with which we will take into account the Pauli principle. The quantity $F(k)/\mathrm{Z}$ is the Fourier component of the distribution density of atomic electrons, which is normalized to unity. For example, the Fourier component of the electron distribution density of a fluorine atom has the form

$$F(k)=\int n(\mathbf{r})\cdot\exp(-i\mathbf{kr})\,d^3\mathbf{r}, \tag{10}$$

where $n(\mathbf{r})=2\left|\psi_{1S}(\mathbf{r})\right|^2+2\left|\psi_{2S}(\mathbf{r})\right|^2+5\left|\psi_{2P}(\mathbf{r})\right|^2$ – electron density of a fluorine atom in the approximation of hydrogen-like wave functions, which approximate the solution of the Hartree-Fock equation for an isolated fluorine atom [6]. Then

$$U_P(k,\sigma)=\frac{4\pi Z_1 Z_2 e^2}{k^2}\left[1-\frac{F_1(k)}{Z_1}\right]^2\left[1-\frac{F_2(k)}{Z_2}\right]^2\exp\left[-\mathrm{k}^2\sigma\right]. \tag{11}$$

The expression for the potential energy of interaction of two atoms (electronic term), taking into account the Pauli principle and plasma oscillations of atomic electrons, has the form

$$U(r)=\int U_P(k,\sigma)\exp(i\mathbf{kr})\frac{\mathrm{d}^3\mathbf{k}}{(2\pi)^3}. \tag{12}$$

The condition for the applicability of the correction in the first order of the perturbation theory to the energy of the system in the unperturbed state has the form

$$|U(r)|\ll\left|E^0\right|,$$

where $E^0=E_1^0+E_2^0$; the energies $E_i^0$ are also presented in [6] together with hydrogen-like wave functions that approximate the solution of the Hartree-Fock equation for an isolated atom.

It is seen that taking into account the plasma oscillations of atomic electrons leads to a self-consistent system of equations

$$\begin{aligned} U'(\mathrm{r}_{\min})&=0\\ U''(\mathrm{r}_{\min})&=\frac{\hbar^2}{8m_e\sigma^4}, \end{aligned} \tag{13}$$

where $U(\mathrm{r})$ depends on $\sigma$ according to (12).

## CALCULATION RESULTS

The numerical solution of the system of equations (13) is shown in Fig. 1 for nitrogen and fluorine molecules. It can be seen that there are points of intersection of the graph of the hyperbola $U''(\mathrm{r}_{\min}) = \hbar^2 / 8m_e\sigma^4$ with the calculated values of the second derivative of the potential energy $U(r)$ at the point of a possible extremum $r_{\min}$ at different values of $U_{rr}(r_{\min})$ for nitrogen and fluorine molecules. It can be seen that there are no intersection points if the potential of isolated nitrogen and fluorine atoms is chosen in the Moliere approximation. The $U_{rr}(r_{\min})$ values at the intersection points in Fig. 1 for nitrogen and fluorine molecules are presented in the table. The graph of the potential energy of interaction of two nitrogen and fluorine atoms depending on the distance between them for two intersection points was calculated using [9] in Fig. 2-Fig. 5. We find the total energy of two isolated nitrogen atoms $E^0_{N_2} \approx -108.8$ a.u. and fluorine $E^0_{F_2} \approx -198.8$ a.u. using [6] (1a.u.=1hartree=27.21eV).

## DISCUSSION OF CALCULATION RESULTS

It is very difficult to compare the results of calculations of the total energy of nitrogen and fluorine molecules with the calculations that were recently published in [1], since the values of the total energies of isolated nitrogen and fluorine atoms in [1] differ significantly from those given in [6]. Since the total energy of an isolated atom is the eigenvalue of the energy of the corresponding eigenwave function, its change should be accompanied by a change in the wave function, which will lead to a change in the atomic form factor, which determines the dependence of the potential energy of interaction between atoms in a molecule. Nevertheless, the three main spectrometric parameters of the interaction potential are in satisfactory agreement with the tabular data [10]. The discrepancy is observed for the fluorine molecule at the intersection point №1.

## ACKNOWLEDGMENTS

The reported study was funded by RFBR, project number 20-07-00236 a.

**List of figures**

**Fig. 1**. The plot of the second derivative of the potential energy $U(r)$ at the point of a possible extremum $r_{\min}$ at different values of $\sigma$ for nitrogen molecules for the Moliere (blue dashed line) and Hartree-Fock (blue solid line) approximation; for fluorine for the approximation of Moliere (red dashed line) and Hartree-Fock (red solid line). The solid black line denotes a hyperbola $U''(\mathrm{r}_{\min}) = \hbar^2 / 8m_e\sigma^4$. ▲– hyperbola intersection points for fluorine, ■ marked points of intersection - for nitrogen.

**Fig. 2**. Potential energy of interaction of two fluorine atoms depending on the distance between them: a) $r/a_0 \in [0;5]$, b) $r/a_0 \in [2;5]$, c) $r/a_0 \in [4.5;9]$. The calculation is performed for the first intersection point.

**Fig. 3**. Potential energy of interaction of two fluorine atoms depending on the distance between them: a) $r/a_0 \in [0;5]$, b) $r/a_0 \in [3.5;6.5]$, c) $r/a_0 \in [6;11]$. The calculation is performed for the second intersection point.

**Fig. 4**. Potential energy of interaction of two nitrogen atoms depending on the distance between them: a) $r/a_0 \in [0;5]$, b) $r/a_0 \in [3;6.5]$, c) $r/a_0 \in [5;11]$. The calculation is performed for the first intersection point.

**Fig. 5**. Potential energy of interaction of two nitrogen atoms depending on the distance between them: a) $r/a_0 \in [0;5]$, b) $r/a_0 \in [3.5;6.5]$, c) $r/a_0 \in [5.5;11]$. The calculation is performed for the second intersection point.

**Table**

| Atom | Point No. | $r_{min}/a_0$ | $U''(r_{min})$, eV/Ang$^2$ | $U(r_{min})$, eV |
|---|---|---|---|---|
| Nitrogen | 1 | 2,034221 | 177,9077117 | -9,98107 |
| | 2 | 2,361961 | 56,27698134 | -4,12273 |
| Fluorine | 1 | 1,341752 | 2599,559918 | -69,4299 |
| | 2 | 2,419894 | 27,17673494 | -2,04894 |

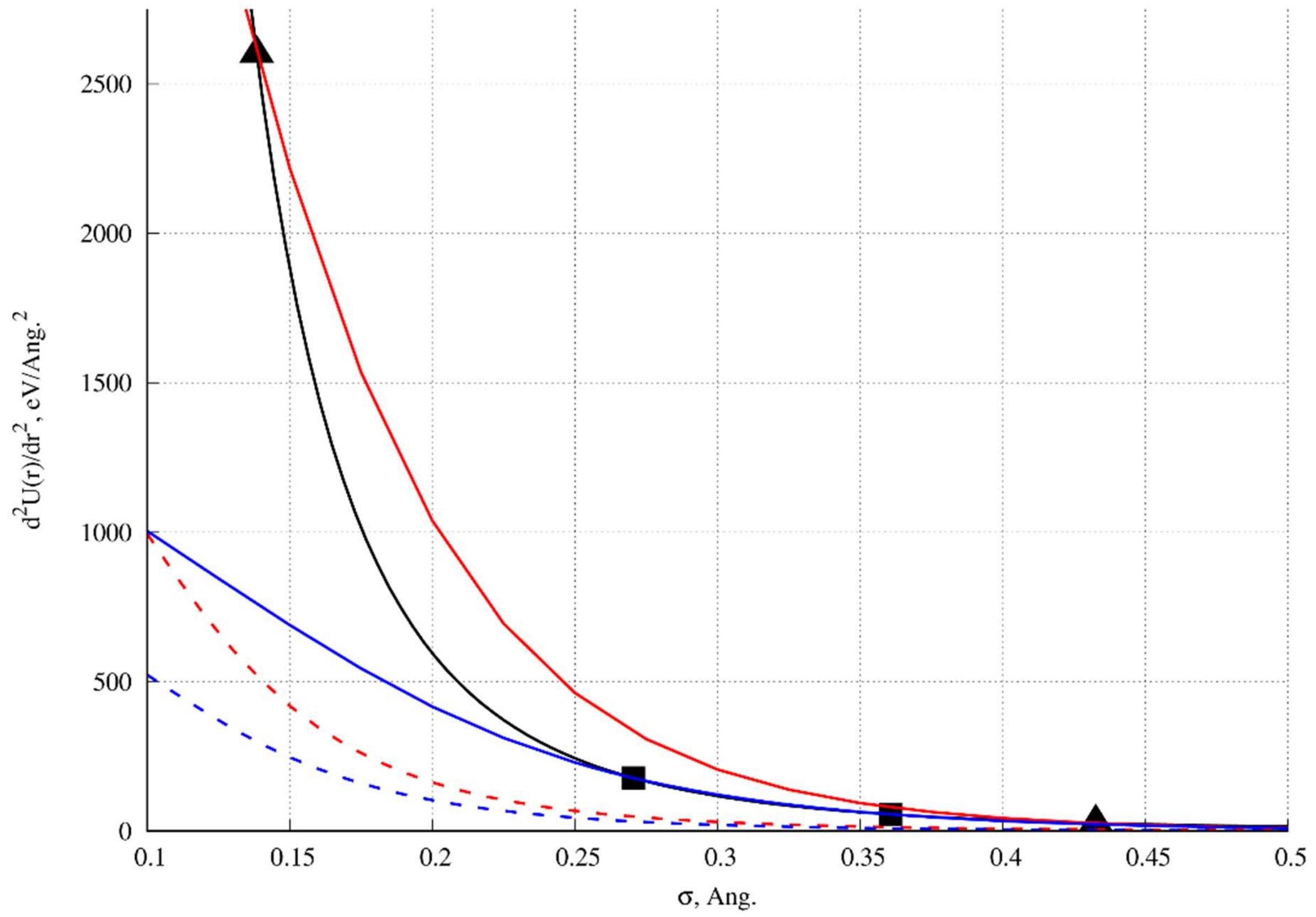

**Figure 1**

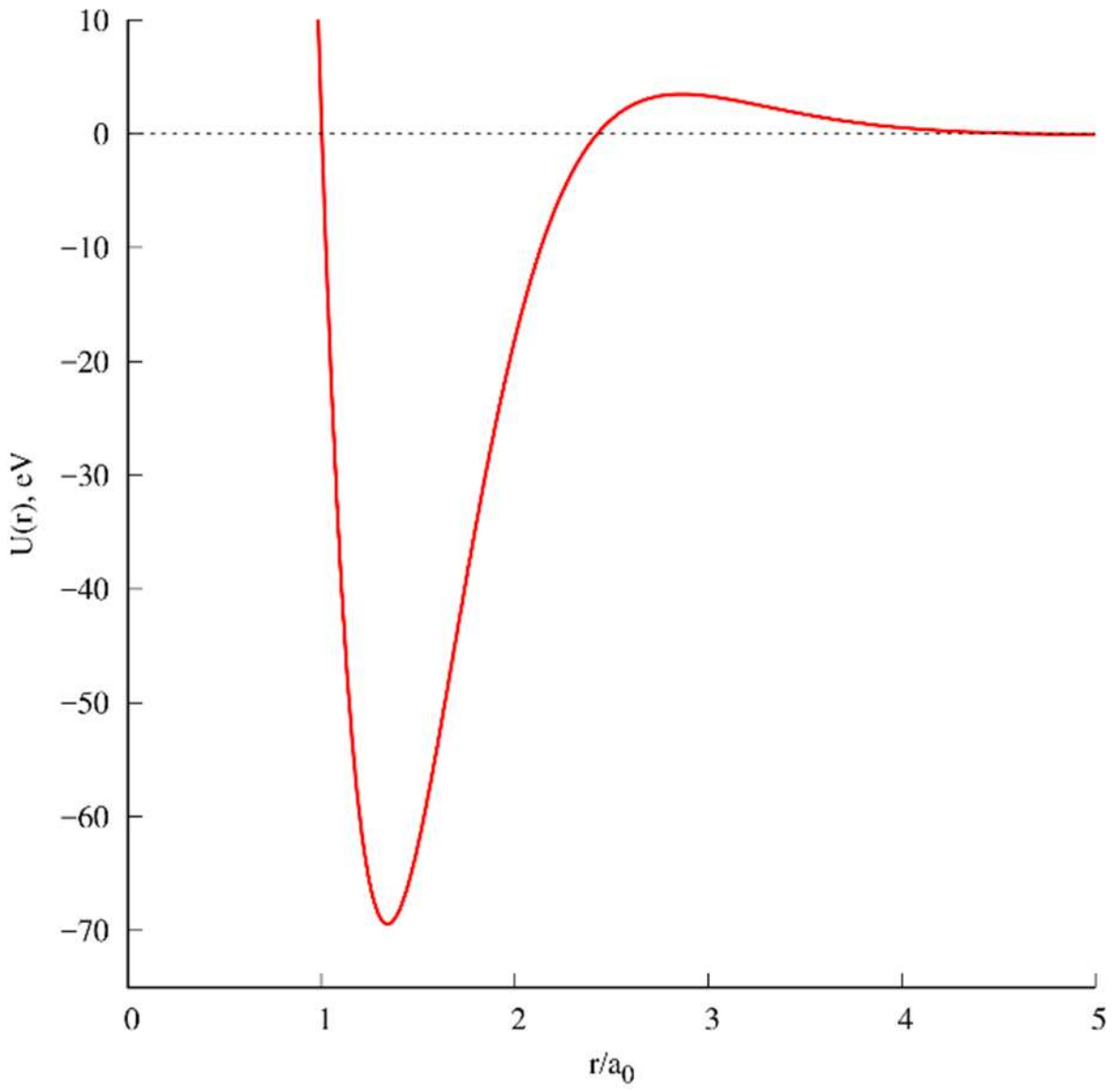

a)

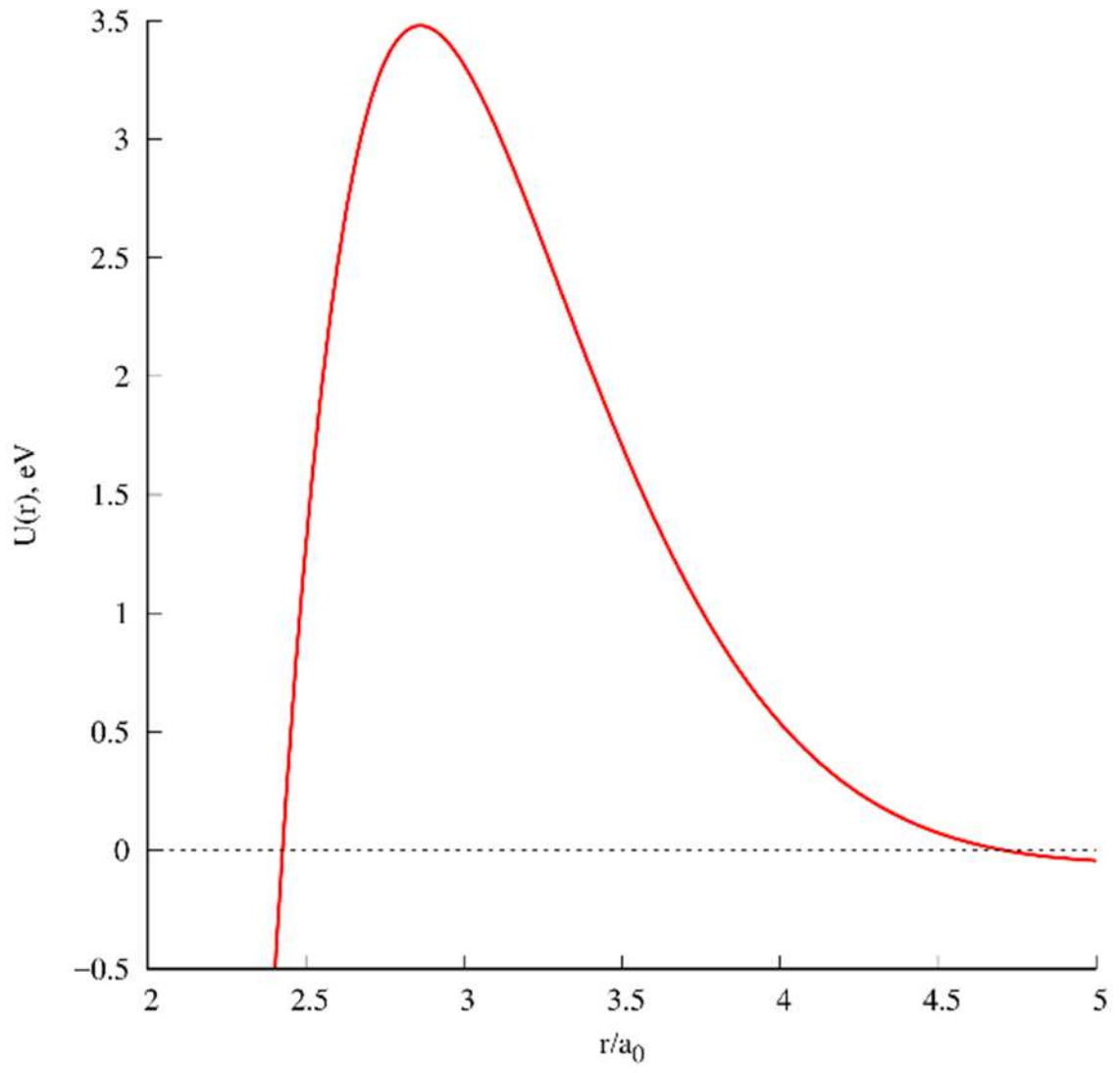

b)

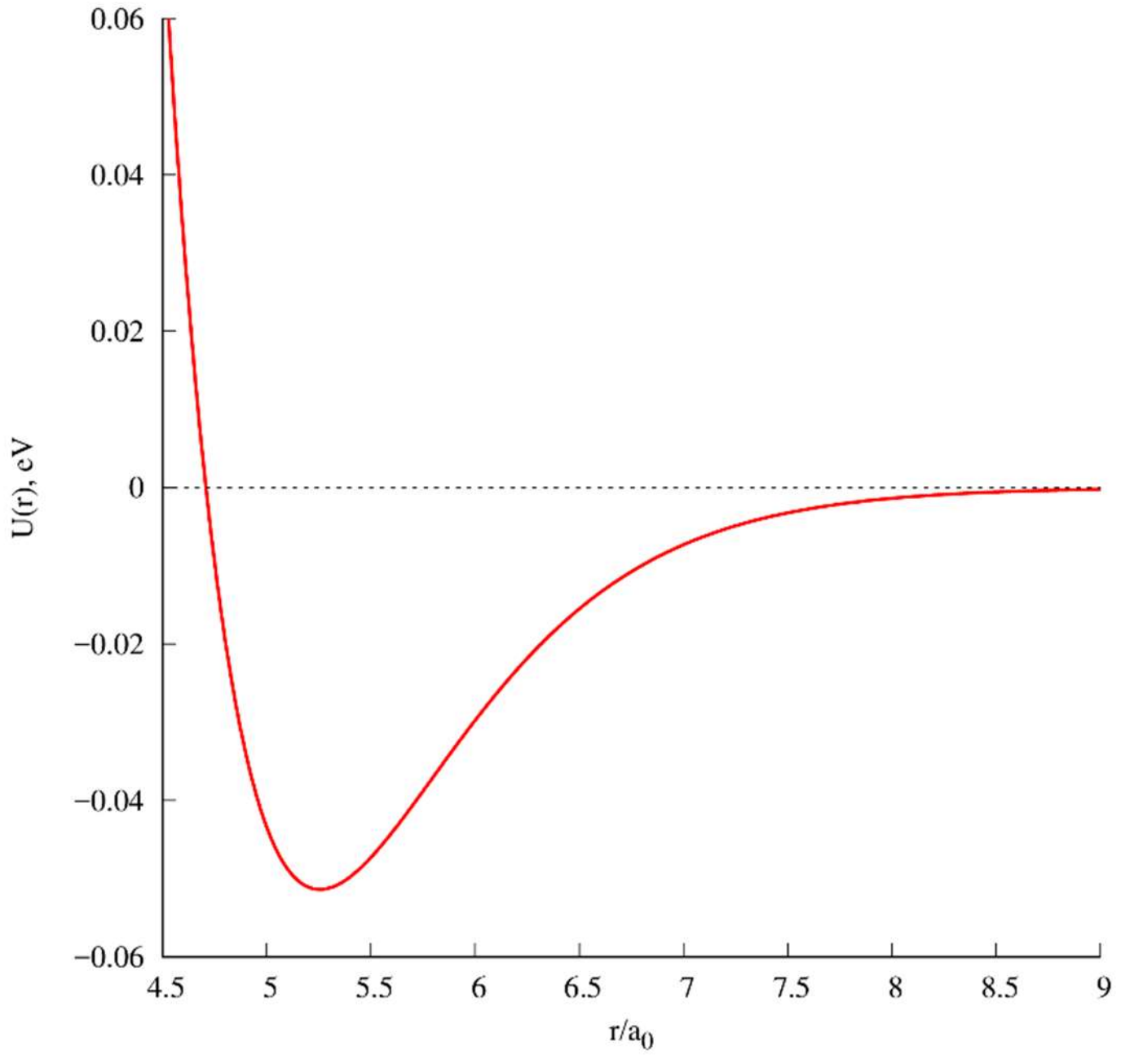

c)

**Figure 2**

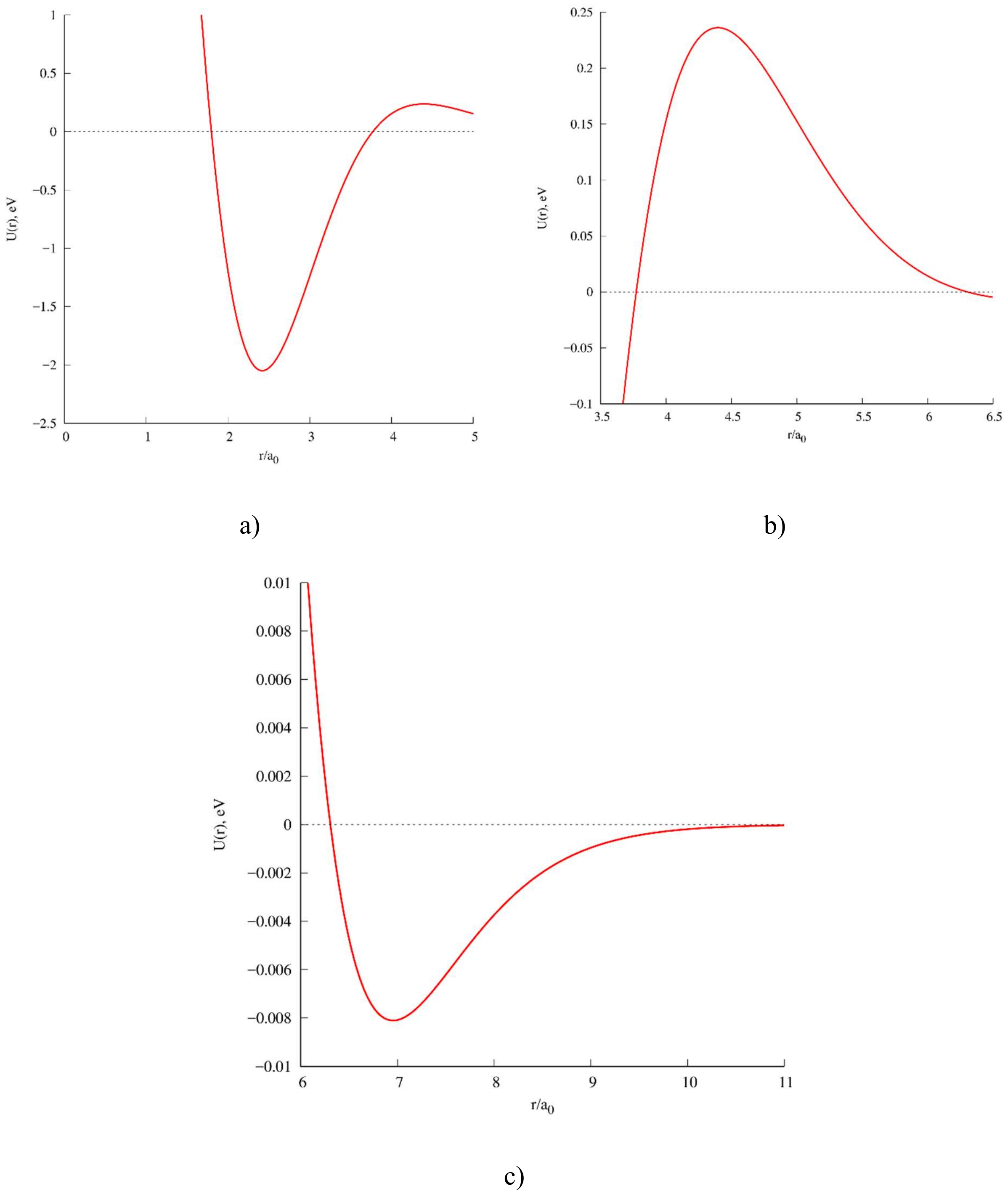

**Figure 3**

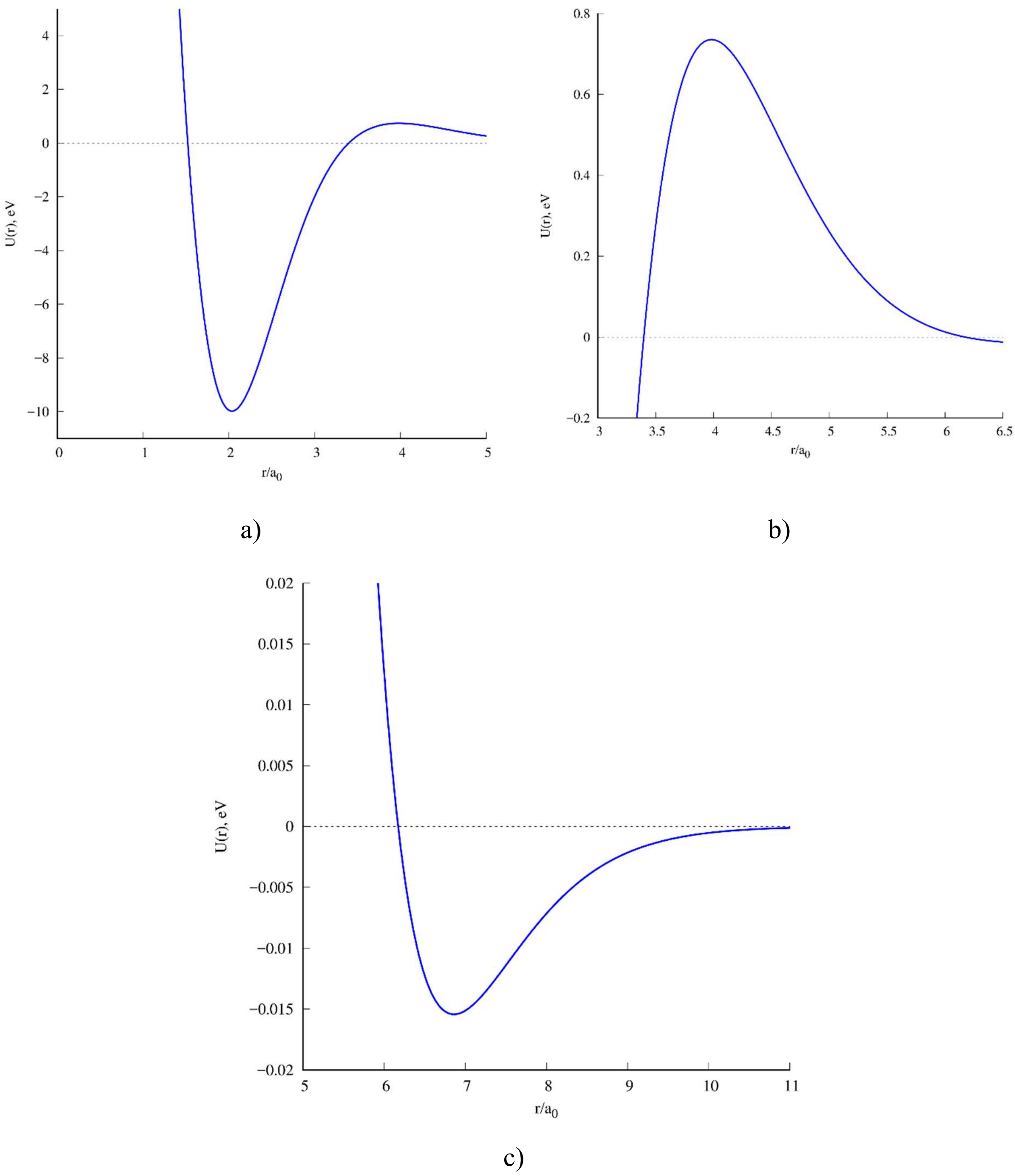

**Figure 4**

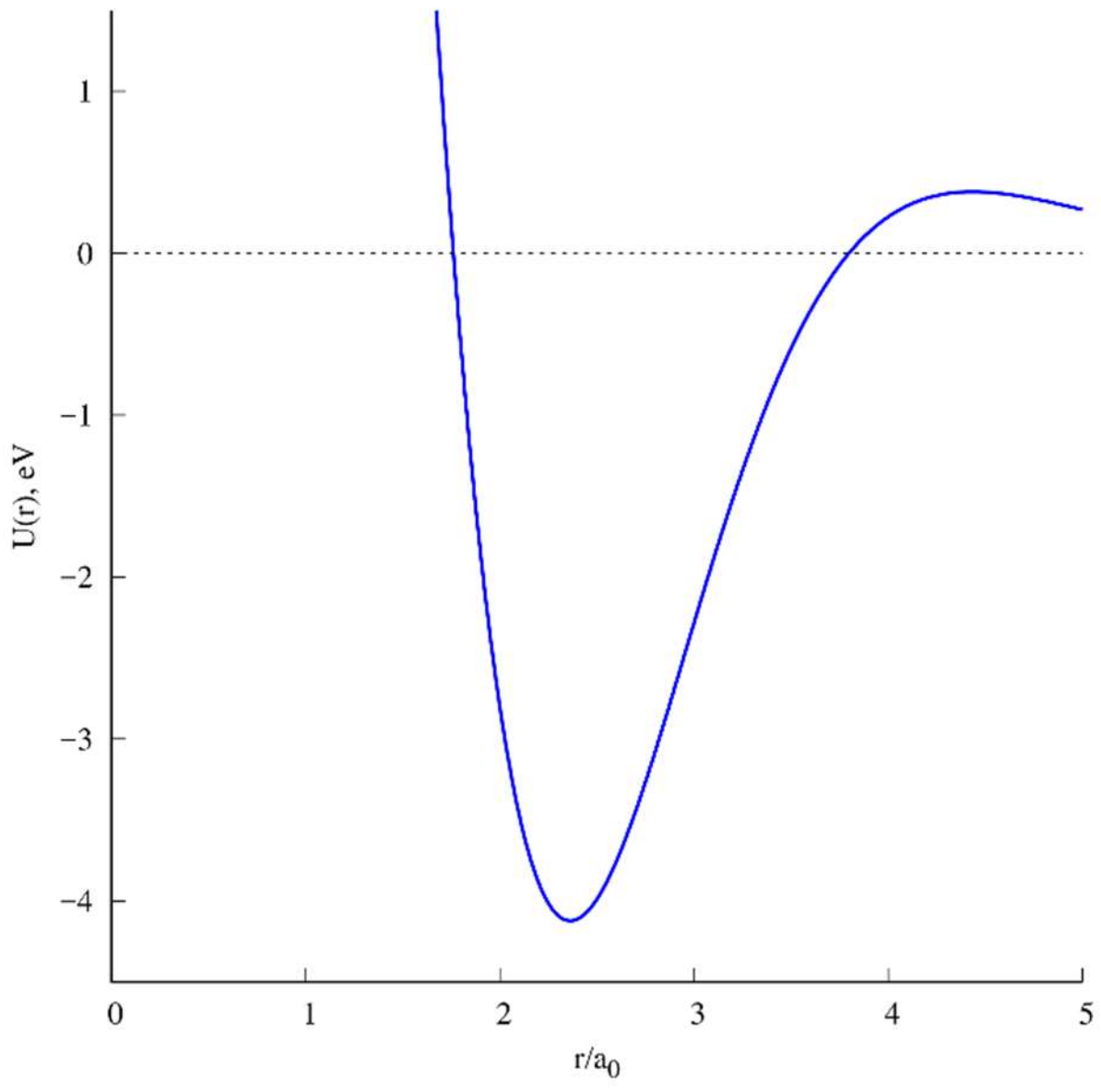

a)

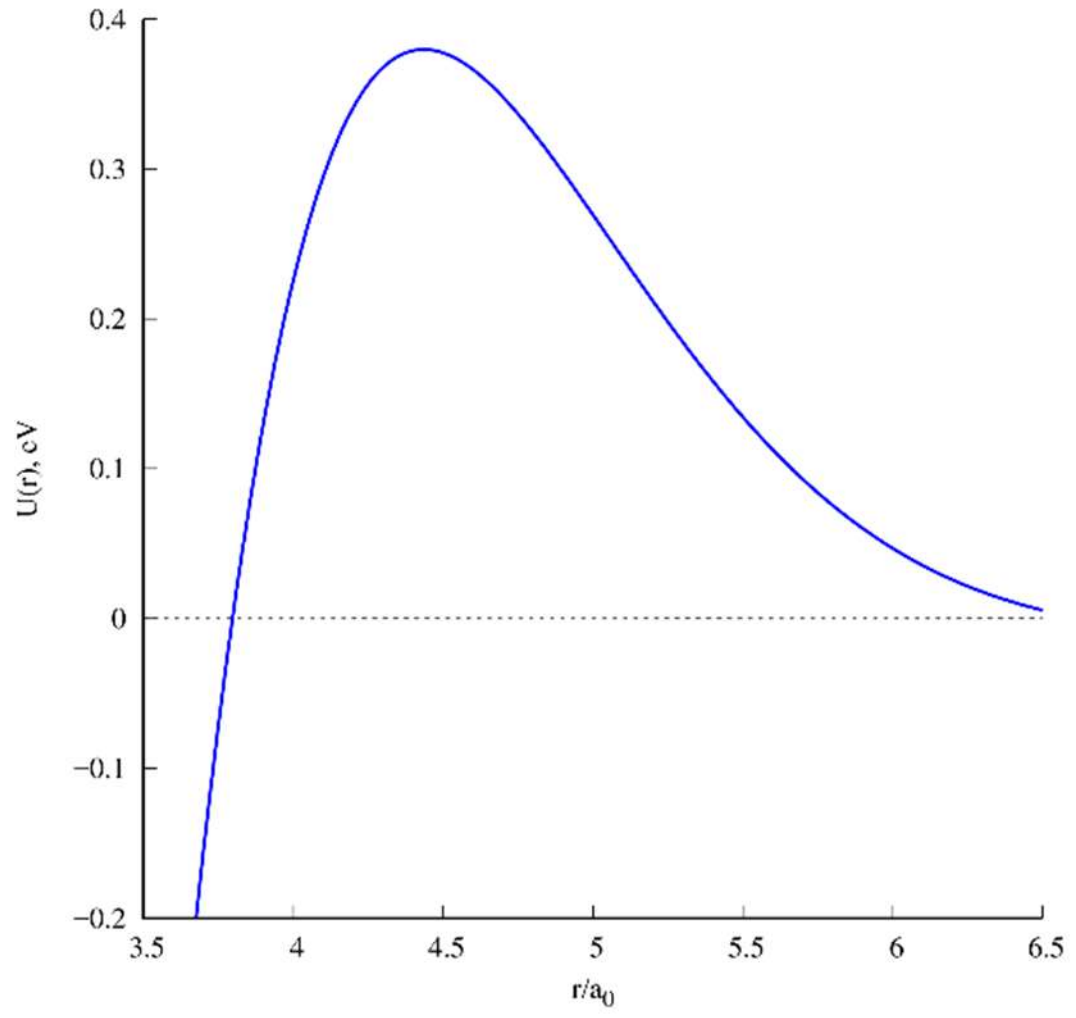

b)

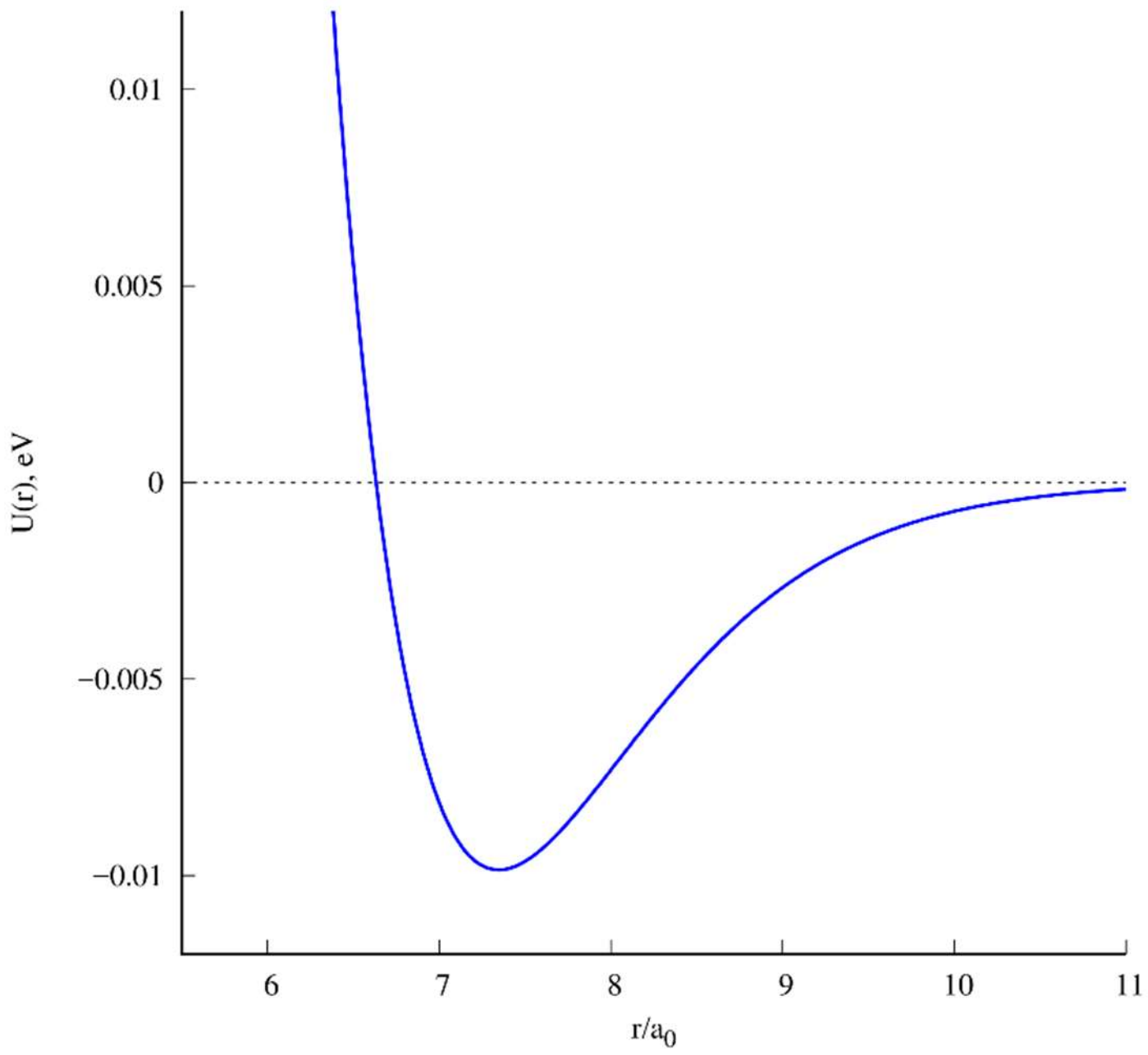

c)

**Figure 5**